# Critical Current of a Granular d-wave superconductor

E. Afsaneh and H. Yavari[1]
Department of Physics University of Isfahan, Isfahan, Iran, 81746

**Abstract:** The twist angle $\gamma$ dependence of the Josephson critical current of d-wave superconductors in one junction and a granular system is considered. Our results show that the twist angle $\gamma$ dependence of the d-wave Josephson critical current is the same for one junction and a granular system. The magnetic field dependences of the critical-current of a granular d-wave superconductor has also been determined by considering the rectangular and circular junction model of an array of small superconducting particles which interacting by Josephson coupling through insulating barriers. We will show that in the case of circular model, the critical current of the Josephson current is larger than that of rectangular one.

## 1-Introduction

Granular superconductors are usually described as a random network of superconducting grains coupled by Josephson weak links [1, 2]. In the high-temperature superconductors (HTSc) ceramics, several experimental groups have found a paramagnetic Meissner effect (PME) at low magnetic fields [3]. Sigrist and Rice [4] proposed that this effect could be a consequence of the intrinsic unconventional pairing symmetry of the HTSc of $d_{x^2-y^2}$ type [5]. Important information about the symmetry of superconducting pairing can be obtained from the measurements of both the dc Josephson effect and the quasiparticle current in tunnel junctions between two HTSc[6]. The dc Josephson effect in unconventional superconductors has been discussed by Geshkenbein et al. [7] and by Sigrist and Rice [4] who demonstrated that the d-wave symmetry of the order parameter may lead to the sign inversion of the Josephson critical current for certain crystal orientations. Under these conditions, the tunnel junction becomes the so-called $\pi$ junction [8]. These results in combination with the paramagnetic behavior of granular HTSc compounds [9] and its theoretical interpretation [4, 10, and 11] serve as a serious argument in favor of d-wave pairing symmetry in HTSc.

Kawamura [12] proposed that a novel thermodynamic phase may occur in zero external magnetic field in unconventional superconductors. This phase is characterized by a broken time-reversal symmetry and is called the chiral glass phase. The frustration effect due to the random distribution of $\pi$ junctions leads to a glass state of quenched in chiralities, which are local loop superconductors circulating over grains and carrying a half-quantum of flux [13].

[1]Corresponding author: Fax:+983117922409
E-mail address: h.yavary@sci.ui.ac.ir

Measurement of the local superconducting gap on single-crystal $Bi_2Sr_2CaCu_2O_{8\_x}$ have shown that, for the superconductivity is not established uniformly. Instead, the system is better described by islands (~50Åwide) having well-established *d*-wave superconductivity, which are separated by regions where no superconducting coherence peaks are present. The global superconductivity is then presumably established due to Josephson tunneling between these islands [14, 15 and 16].

It is clear that an array with sufficiently strongly coupled grains should be able to maintain the superconducting coherence in the whole sample because the coupling reduces the phase fluctuations [17]. In contrast, in the opposite limit of weak coupling, one expects that the strong Coulomb interaction should lead to the Coulomb blockade of the Cooper pairs in analogy with the Coulomb blockade of electrons in granular metals in the low-coupling regime.

In order to quantify this intuitive statement [18], following the earlier idea of [19], suggested comparing the energy of the Josephson coupling of neighboring grains with the Coulomb energy. Indeed, the Josephson coupling tends to lock the phases of neighboring grains and to delocalize the Cooper pairs, while the Coulomb interaction tends to localize the Cooper pairs and thus enhance the quantum phase fluctuations. Comparing the Josephson energy $E_j = \frac{\pi g \Delta}{2}$, with the Coulomb energy one comes to the conclusion that samples with $g > g_s \sim \frac{E_C}{\Delta}$ should be superconductors, while those with $g < g_s$ are insulators [17].

The Josephson energy $E_J$ is finite in the zero-temperature limit, whereas the charging energy $E_C = \frac{e^2}{2C_T}$ ($C_T$ is the total Capacitance) is strongly suppressed. At low temperatures, the number (and charge) fluctuations which are inherent to a superconducting state have negligible energy cost in spite of the small size of superconducting islands. Since $E_C \ll E_J$ the system is robust against destruction of the superconducting order. So it can be deduced that a system with granular SC regions will be more stable in the case of a d-wave symmetry compared to an s-wave case [20].

This result was later derived with a superconducting granular array model including both Coulomb and Josephson interactions in a number of theoretical works [21, 22 and 23] using different methods.

We consider granular samples that are relatively good metals in their normal state, such that $g \gg 1$. This allows us to neglect the effect of the suppression of the critical temperature by the Coulomb interaction and fluctuations. In this region( $g \gg 1$ ), all effects of the weak localization and the charging effects have to be small which would imply that the resistivity could not considerably depend on the magnetic field.

In this Letter, we focus on the nature of the Josephson coupling in granular $d$-wave superconductors. This case is qualitatively different from the granular $s$-wave case in which there are no low-energy quasiparticles and which has been investigated extensively [24]. In contrast to the $s$-wave case, the $d$-wave gap function $\Delta(\boldsymbol{p}) = \Delta_0[\cos(p_x) - \cos(p_y)]$ has no minimal value. The nodal points on the Fermi surface produce cooper pairs with an arbitrarily small energy gap.

D. Dominguez et al. [25] studied the effects of an electric field in the transport of bulk granular superconductors with different kinds of disorder. They found that for a d-wave granular superconductor with random junctions the critical current always increases after applying a strong magnetic field, regardless of the polarity of the field. Also the influence of magnetic field on the superconducting transition in granular (Bi, Pb)–Sr–Ca–Cu–O superconductors was studied by M. Gazda [26, 27].

In this paper, first, the twist angle $\gamma$ dependence of the Josephson critical current of one junction and a granular d-wave superconductor is considered and then the magnetic field dependences of the critical-current of a granular d-wave superconductor is determined.

**2- Formalism**

We consider two superconducting grains, indexed by $\alpha = L, R$ each of which has a well-established d-wave order parameter. The Hamiltonian of the systems is given by [28]

$$H = H_L + H_R + H_T + H_Q \tag{1}$$

where, $H_{L(R)}$ is the Hamiltonian of the left and right hand side of the junction which contains a single-particle kinetic term $H$ and an effective local attraction $H_{BCS}$ that leads to d-wave superconductivity

$$H_{L(R)} = \sum_{\sigma,\acute{\sigma}} \int d^3r \, \psi^\dagger_{R(L)\sigma}(\vec{r}) \left( -\frac{\hbar^2 \nabla^2}{2m^*} - \mu \right) \psi_{R(L)\sigma}(\vec{r})$$

$$-\frac{1}{2}\sum_{\sigma,\acute{\sigma}} \int d\vec{r}d\vec{r}'\psi^{\dagger}_{R(L)\sigma}(\vec{r})\psi^{\dagger}_{R(L)\sigma'}(\vec{r}') g_{R(L)}(\vec{r}-\vec{r}')\psi_{R(L)\sigma'}(\vec{r}')\psi_{R(L)\sigma}(\vec{r}) \qquad (2)$$

where $m$ is the electron mass, $\mu$ is the chemical potential, and $\psi(\psi^{\dagger})$ is the fermion field operator. In order to obtain the anisotropic order parameter, the anisotropic attractive interaction $g(\vec{r}-\vec{r}')$ has to be taken into account. The third term in Hamiltonian (Eq. (1)), i.e.,

$$H_t = \sum_{\sigma} \int d\vec{r}d\vec{r}' t(\vec{r}-\vec{r}')\psi^{\dagger}_{R\sigma}(\vec{r}) \psi_{L\sigma}(\vec{r}') + h.c \qquad (3)$$

describes the tunneling of electrons between the two sides of the junctions, where $t(\vec{r}-\vec{r}')$ is the probability amplitude for an electron to tunnel from position $\vec{r}$ in one grain to position $\vec{r}'$ in the another grain, and

$$H_Q = \frac{(Q_R-Q_L)^2}{8C} \qquad (4)$$

is the charging Hamiltonian, where $C$ is the capacitance of the junction and $Q_{R(L)}$ is the operator for the charge on the grain $R(L)$, which can be written as

$$Q_{R(L)} = e\sum_{\sigma} \int d\vec{r}\psi^{\dagger}_{R(L)}(\vec{r}) \psi_{R(L)}(\vec{r}) \qquad (5)$$

We neglect Coulomb interaction because it is well screened in the optimally doped regime [28]. The tunneling current through the junction is expressed as the rate of change of the number of particles on, for example, the left-hand side of the junction $N_L = \sum_{p,\sigma} c^{\dagger}_{p\sigma}c_{p\sigma}$. The total current through the tunneling interface is defined as the average value of this operator

$$I(t) = -e\langle \dot{N}_L(t)\rangle \qquad (6)$$

where

$$\dot{N}_L = \frac{i}{\hbar}[H, N_L] = \frac{i}{\hbar}[H_t, N_L] \qquad (7)$$

From Eqs. (6) and (7) we have

$$I(t) = -\frac{ei^2}{\hbar}\int_{-\infty}^{t} dt' \left\langle \left\{ \sum_{kp}[T_{kp}e^{-ieVt}c^{\dagger}_k(t)c_p(t) - T^*_{kp}e^{ieVt}c^{\dagger}_p(t)c_k(t)], \right.\right.$$

$$\left.\left.\sum_{k'p'}[T_{k'p'}e^{-ieVt}c^{\dagger}_{k'}(t')c_{p'}(t') - T^*_{k'p'}e^{ieVt}c^{\dagger}_{p'}(t')c_{k'}(t')]\right\}\right\rangle \qquad (8)$$

where $V = \frac{\mu_R - \mu_L}{e}$ is the applied voltage ($\mu_i$ is the chemical potential of the grains).

Using Matsubara Green's functions techniques, Eq. (8) can be written as

$$I = \frac{2e}{\hbar}\text{Im}\left[\sum_{kp}|T_{k,p}|^2 \frac{1}{\beta}\sum_{ip_n} \mathcal{G}_L(\boldsymbol{p}, ip)\mathcal{G}_R(\boldsymbol{k}, ip-i\omega)\right.$$

$$+\frac{1}{\beta}\Sigma_{kp,ip_n} T_{k,p}T_{-k,-p}\mathcal{F}_L^\dagger(\boldsymbol{k},ip)\mathcal{F}_L(\boldsymbol{p},ip-i\omega)e^{i[\varphi-2eV]}\Big] \quad (9)$$

here the diagonal $\mathcal{G}$ and off-diagonal $\mathcal{F}$ components of the Matsubara Green functions are given by

$$\mathcal{G}(\boldsymbol{k},i\omega_n) = \frac{1}{i\omega_n - E(\boldsymbol{k})} \quad (10)$$

$$\mathcal{F}(\boldsymbol{k},i\omega_n) = \frac{\Delta(\boldsymbol{k})}{i\omega_n - E(\boldsymbol{k})} \quad (11)$$

From Eqs. (9), (10), and (11) we have

$$I = \frac{2e}{\hbar}\text{Im}\sum_{kp}\int\frac{d\omega}{(2\pi)^3}\int\frac{d\omega'}{(2\pi)^3}[f_L(\omega) - f_R(\omega')]$$

$$\times\left\{|T_{k,p}|^2\frac{A(\boldsymbol{k},\omega)A(\boldsymbol{p},\omega)}{\omega-\omega'+i\eta} + T_{k,p}T_{-k,-p}\frac{B(\boldsymbol{k},\omega)B(\boldsymbol{p},\omega)}{\omega-\omega'+i\eta}e^{i[\varphi-2eV]}\right\} \quad (12)$$

where $A$ and $B$ are the spectral functions of the normal and anomalous Greens functions, $\varphi = \varphi_L - \varphi_R$ is the phase difference between two grains, and $f_{L(R)}$ the Fermi distribution at the chemical potential $\mu_{L(R)}$. The $T_{k,p}$ is the matrix elements that transfer electrons from a state $\boldsymbol{k}$ in Left grain to a state $\boldsymbol{p}$ in Right grains, which is a momentum-dependent tunneling matrix element across the junction[20]. We consider a simple model of tunneling, $|T_{k,p}|^2 = |T_0|^2 + |T_1|^2\delta(\boldsymbol{k}-\boldsymbol{p})$. This model allows us to capture both the momentum nonconserving ($|T_0|^2$) and the momentum conserving ($|T_1|^2$) tunneling processes in a compact form.

In the experimental samples, since the underlying material is a single crystal, we can assume that the superconducting gaps on the different islands have the same relative orientation (see Fig. 1 b). However, for a $d$-wave case, it is necessary to include a momentum conserving term $|T_1|^2$ to get a nonzero Josephson coupling between the grains. This term represents tunneling over an extended area instead of a point contact.

However, our results are also applicable to artificially created Josephson contacts between large $d$-wave superconducting samples.

The first term in Eq. (12) describes the quasiparticle current $I_Q$ and the second term the supercurrent $I_J$. The spectral densities have the form

$$A(\boldsymbol{k},\omega) = \pi\left[\left(1 + \frac{\varepsilon_k}{E(\boldsymbol{k})}\right)\delta(\omega - E(\boldsymbol{k})) + \left(1 - \frac{\varepsilon_k}{E(\boldsymbol{k})}\right)\delta(\omega + E(\boldsymbol{k}))\right] \quad (13)$$

$$B(\boldsymbol{k},\omega) = \pi\frac{\Delta(\boldsymbol{k})}{E(\boldsymbol{k})}[\delta(\omega + E(\boldsymbol{k})) - \delta(\omega - E(\boldsymbol{k}))] \quad (14)$$

where $E(\mathbf{k}) = (\varepsilon_\mathbf{k}^2 + \Delta(\mathbf{k})^2)^{\frac{1}{2}}$.

After transform the summation over $\mathbf{k}$ and $\mathbf{q}$ to the integral as

$$\sum_\mathbf{k} = \int \frac{d^3k}{(2\pi)^3} = \frac{1}{(2\pi)^3} \int dk k^2 \int d\theta \sin\theta \int d\varphi = N(0) \int \frac{d\Omega}{4\pi} \tag{15}$$

where $N(0)$ is the density of state at the Fermi energy, also from now by above description we show $|T_{k,p}|^2$ with $<|T|^2>$ symble. In the following we rewrite the Eq. (12) as the summation of quasiparticle $I_Q$ and Josephson $I_J$ currents ($\omega_0 = eV_0/\hbar$)

$$I_Q(T) = \frac{8e}{\hbar} N_L(0) N_R(0) <|T|^2> P \int_0^{+\infty} d\omega \int_0^{+\infty} d\acute\omega \int \frac{d\Omega}{4\pi} \int \frac{d\acute\Omega}{4\pi} \frac{A_L(\omega) A_R(\acute\omega)}{\omega - \acute\omega - \omega_0} [f(\omega) - f(\acute\omega)] \tag{16}$$

$$I_J(T) = \frac{8e}{\hbar} N_L(0) N_R(0) <|T|^2> P \int_0^{+\infty} d\omega \int_0^{+\infty} d\acute\omega \int \frac{d\Omega}{4\pi} \int \frac{d\acute\Omega}{4\pi} \frac{B_L(\omega) B_R(\acute\omega)}{\omega - \acute\omega - \omega_0} [f(\omega) - f(\acute\omega)] \tag{17}$$

here, $P$ indicates the principal part of the integral.

We restrict ourselves to calculate the Josephson current for $V=0$. By using Eq. (14) for $B_i(\omega)$ in to Eq. (17) we get

$$I_J(T) = \frac{2}{\pi R_N} P \int_0^{+\infty} d\omega \int_0^{+\infty} d\acute\omega \int \frac{d\Omega}{4\pi} \int \frac{d\acute\Omega}{4\pi} \frac{\Delta_L(\Omega)}{\sqrt{\omega^2 - \Delta_L^2(\Omega)}} \frac{\Delta_R(\acute\Omega)}{\sqrt{\omega^2 - \Delta_R^2(\acute\Omega)}} \left[ \frac{1}{\omega + \acute\omega} + \frac{2\acute\omega f(\omega)}{\omega^2 - \acute\omega^2} - \frac{2\omega f(\acute\omega)}{\omega^2 - \acute\omega^2} \right] \tag{18}$$

where $R_N \equiv \frac{\hbar}{4e N_R(0) N_L(0) <|T|^2>}$.

After some mathematical operations and changing the variables, the Josephson current at $T = 0$ becomes

$$I_J(T = 0) = \frac{2}{R_N} \int \frac{d\Omega}{4\pi} \int \frac{d\acute\Omega}{4\pi} \frac{\Delta_L(\Omega) \Delta_R(\acute\Omega)}{\Delta_L(\Omega) + \Delta_R(\acute\Omega)} K\left(\frac{|\Delta_L(\Omega) - \Delta_R(\acute\Omega)|}{\Delta_L(\Omega) + \Delta_R(\acute\Omega)}\right) \tag{19}$$

where $K(x)$ is the complete elliptic integral of the first kind. Using the asymptotic expression $K(x)_{x \to 0} \cong \frac{\pi}{2}$, for $\Delta_L \cong \Delta_R$ Eq. (19) gives

$$I_J(T = 0) = \frac{\pi}{R_N} \int \frac{d\Omega}{4\pi} \int \frac{d\acute\Omega}{4\pi} \frac{\Delta_L(\Omega) \Delta_R(\acute\Omega)}{\Delta_L(\Omega) + \Delta_R(\acute\Omega)} \tag{20}$$

According to Fig. 1 (a) the Josephson current depends on the twist angle $\gamma$. Using the energy gap parameter for the left $\Delta_L(\Omega) = \Delta_0 \cos 2\theta$ and right $\Delta_R(\Omega) = \Delta_0 \cos 2(\theta + \gamma)$ hand side of the junction, we can write

$$I_J(\gamma) = \frac{\Delta_0}{2R_N} \int d\theta \frac{\cos 2\theta \cos 2(\theta + \gamma)}{\cos 2\theta + \cos 2(\theta + \gamma)} \tag{21}$$

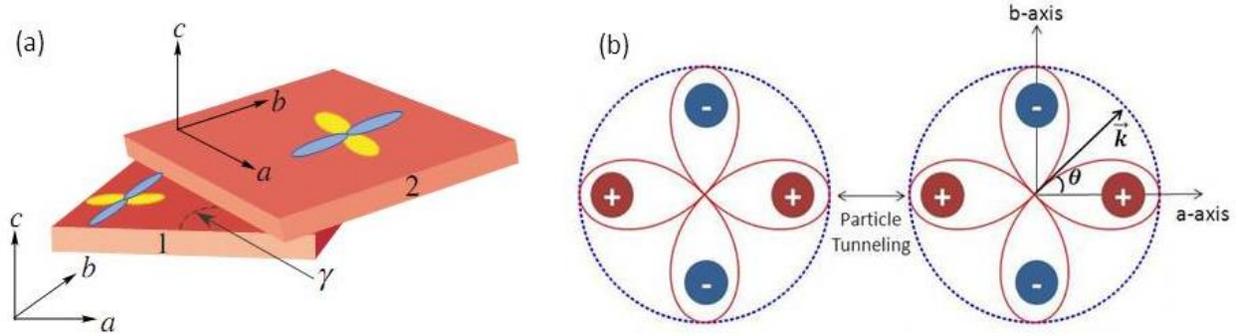

Figure 1 (a): Schematic picture of the c-axis twist Josephson junction of d-wave superconductors. $\gamma$ is the twist angle around the c axis[29], (b): Schematic of two d-wave superconductors grains when the junction is node to node ($\gamma = 0$)

For especial case node-to-node junction ($\gamma = 0$) (Fig.1 (b)) Eq. (21) reduces to the well-known result [22].

$$I_J(0) = \frac{2e}{\hbar} N^2(0) <|T|^2> \qquad (22)$$

Figure 2 shows the twist angle $\gamma$ dependence of the Josephson critical current $I_J(\gamma)/I_J(0)$. Note that for $\gamma > \pi/4$, the sign of the current changes. Thus, in this case the $\pi$ junction is formed [30].

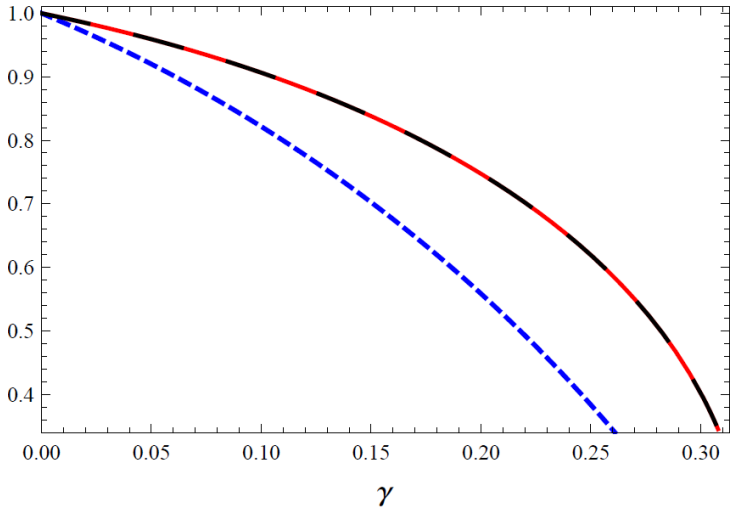

Figure 2: The twist angle $\gamma$ dependence of the Josephson critical current, "solid line for one junction $\frac{I_J(\gamma)}{I_J(0)}$", "dotted line for granular system $\frac{I_{J-g}^2(\gamma)}{I_{J-g}^2(0)}$ and "dashed line for root mean square (rms) granular system $\frac{I_{rms-g}(\gamma)}{I_{rms-g}(0)}$

It is noted that to derive Eq. (21) we have considered c-axis junction. It is interesting to consider ab plane junction. Zero-bias conductance peaks and non-monotonic temperature dependence of the critical current is possible in ab plane tunneling due to the presence of Andreev resonant state specific to d-wave superconductor [31, 32, 33, 34, 35]. Critical Current of a Granular d-wave superconductor in the presence of Andreev resonant state is under our consideration for future works.

In order to analyze the Josephson critical current of granular d-wave superconductor, we consider Meilikhov's model, in which, the weak link between grains is formed in the region of plane segments [36]. In this case, the banks of an intergranular Josephson junction are in the form of a circle whose radius r is proportional to the granule size $r = ka$ ($a$ is the average granule size). Suppose that $d$ is the junction thickness.

To calculate the mean square of Josephson critical current and the root mean square critical current for d-wave granular system (Fig 2), we assume that the randomness of the Josephson lattice is governed by Gauss-like fluctuations of the form

$$P(r) = \frac{32r^2}{\pi^2 k^3 a^3} e^{-\frac{4r^2}{\pi k^2 a^2}} \tag{23}$$

Thus Gaussian averaging leads to

$$\frac{I_{J-g}^2(\gamma)}{I_{J-g}^2(0)} = -\frac{1}{2}\text{Sin}[\gamma]\left\{1 + \left(\text{Log}\left[\text{Cos}\left[\frac{\gamma}{2}\right] - \text{Sin}\left[\frac{\gamma}{2}\right]\right] - \text{Log}\left[\text{Cos}\left[\frac{\gamma}{2}\right] + \text{Sin}\left[\frac{\gamma}{2}\right]\right]\right)\text{Sin}[\gamma]\right\} +$$

$$\frac{1}{2}\left\{\text{Cos}[\frac{21\pi}{200} - \gamma] + \left(\text{Log}\left[\text{Cos}\left[\frac{79\pi}{400} + \frac{\gamma}{2}\right] - \text{Sin}\left[\frac{79\pi}{400} + \frac{\gamma}{2}\right]\right] - \text{Log}\left[\text{Cos}\left[\frac{79\pi}{400} + \frac{\gamma}{2}\right] + \text{Sin}\left[\frac{79\pi}{400} + \frac{\gamma}{2}\right]\right]\right)\text{Sin}[\gamma]^2\right\} \tag{24}$$

Shown in Fig. 2 is a comparison between the critical current of one d-d junction, mean square and root mean square of critical current for granular d-wave superconductor. Three curves have the same behavior with respect $\gamma$, but the $I_{rms}$ of the granular system completely fitted with critical current of one junction.

To describe the magnetic field dependence of d-d junction critical current, the small junction compared with the Josephson penetration depth and node-to-node $\gamma = 0$ junction is used.

We use a coordinate system with the $z$ axis normal to the plane of the junction, and an external magnetic field $H$ applied in the $y$ direction. The magnetic flux is gives by

$$\varphi(x) = \frac{2\pi d}{\phi_0} H_y x + \varphi_0 \tag{25}$$

where, $d = t + \lambda_R + \lambda_L$ ($\lambda_L$ and $\lambda_R$ are the London depths in the two superconductors and $t$ is the dielectric barrier thickness), $\phi_0 = hc/2e$ is the flux quantum and $\varphi_0$ is an integration constant.

Because the phase difference along the junction can vary by virtue of the presence of a magnetic field passing through the junction, so can the critical currents. The total current that flows through the junction is obtained by integrating the local critical current density $\mathcal{J}(x)$ over the junction area $A$.

In general, the total current may be written as

$$I(B) = \left| \int_{-\infty}^{\infty} dx \mathcal{J}(x) e^{i\eta x} \right| \tag{26}$$

where $\mathcal{J}(x) = \int dy\, j_1(x,y)$ ($j_1$ is the critical density of the Josephson current in the junction) and $\eta = 2\pi B d/\phi_0$.

We consider two specific junction geometries: rectangular and circular for d-d junctions (Fig. 3).

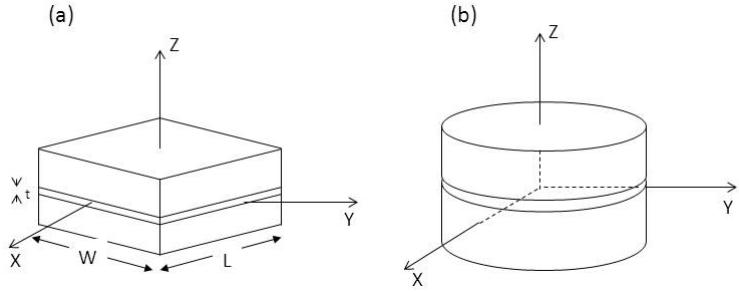

Figure 3 Geometrical configuration: (a) Rectangular geometry (b) Circular geometry

In the case of a rectangular Josephson junction, the critical current becomes [37, 38, 39]

$$I_{Rec}\left(\frac{\phi}{\phi_0}\right) = I_1 \left| \frac{\sin^2\left(\frac{\pi \phi}{2\phi_0}\right)}{\frac{\pi \phi}{2\phi_0}} \right| \tag{27}$$

where $\phi = BLd$, is the magnetic flux through the junction, $I_1 = J_1 WL$, and $\phi_0 = \frac{hc}{2e}$ (Fig. 4(a)).

For circular junction we can write

$$\mathcal{J}(x) = \int_{-\sqrt{R^2-x^2}}^{\sqrt{R^2-x^2}} dy J_1 = 2J_1 \sqrt{R^2 - x^2} \tag{28}$$

where $R$ is the radius of the junction.

To calculate the maximum Josephson current in applied magnetic field we apply the additional phase of $\pi$ in the critical current distribution

$$I_{Cir}(\eta) = 2J_1 \left| \int_{-R}^{0} dx \sqrt{R^2 - x^2}\, e^{i\eta x} + \int_{0}^{R} dx \sqrt{R^2 - x^2}\, e^{i(\eta x + \pi)} \right| \tag{29}$$

where $\eta = \frac{2\pi B d}{\phi_0}$ [40]. Thus, the critical current becomes,

$$I_{Cir}(\eta) = I_1 \left| \frac{H_1(\eta R)}{\frac{1}{2}\eta R} \right| \tag{30}$$

here $I_1 = \pi R^2 J_1$ and $H_1(x)$ is the first order Struve function.

Eq. (30) can be written as

$$I_{Cir}\left(\frac{\phi}{\phi_0}\right) = 2I_1 \left|\frac{H_1\left(\pi\frac{\phi}{\phi_0}\right)}{\pi\frac{\phi}{\phi_0}}\right| \tag{31}$$

with $\phi = 2BRd$ (Fig.4 (b))

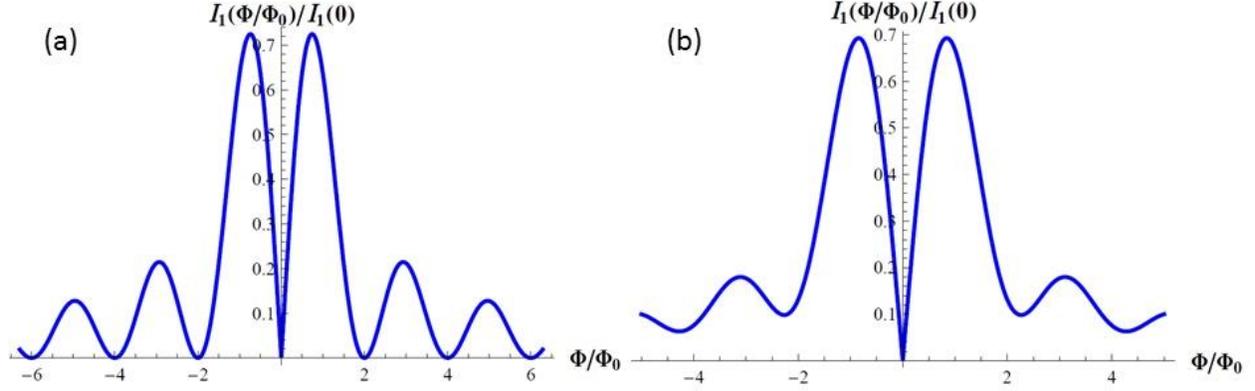

Figure4: Theoretical magnetic field dependence of the maximum Josephson current in d-d junction with (a) Rectangular (b) Circular geometry

Now we discussed the critical current of a granular d-wave superconductor in magnetic field with rectangular (Meilikhov's) and circular model for d-d junctions. For these models Eqs. (27) and (31) respectively reduce to

$$I_{Rec}(B_0) = j_1 \pi r^2 \frac{\sin^2\left(\frac{\pi r d B_0}{2\phi_0}\right)}{\frac{\pi r d B_0}{2\phi_0}} \tag{32}$$

$$I_{Cir}(B_0) = j_1 \pi r^2 \left(\frac{H_1\left(\frac{\pi r d B_0}{\phi_0}\right)}{\frac{\pi r d B_0}{2\phi_0}}\right) \tag{33}$$

We stress that the magnetic field behavior of the critical current is strongly depends on the form of the grain distribution function. For performing the averaging, we restrict ourselves to Gaussian distribution law. Since, it is quite difficult to average the modulus in Eqs. (32) and (33), we calculate the mean square critical current of the junctions. By using Eq. (23), the configurational averaging in Eqs. (32) leads to

$$<I^2_{Rec}(H_a)> = \frac{\pi k^2 a^2 j_1^2 \phi_0^2}{16 d^2 H_a^2}\left\{\frac{9}{4} - \left[3 - 12\frac{\pi^3 k^2 a^2 d^2 H_a^2}{4\phi_0^2} + 4\left(\frac{\pi^3 k^2 a^2 d^2 H_a^2}{4\phi_0^2}\right)^2\right] e^{-\frac{\pi^3 k^2 a^2 d^2 H_a^2}{4\phi_0^2}}\right.$$

$$+\left[\frac{3}{4}-12\frac{\pi^3k^2a^2d^2H_a^2}{4\phi_0^2}+16\left(\frac{\pi^3k^2a^2d^2H_a^2}{4\phi_0^2}\right)^2\right]e^{-\frac{\pi^3k^2a^2d^2H_a^2}{\phi_0^2}}\Bigg\} \quad (34)$$

Eq. (34), can also be written as

$$<I_{Rec}^2(H_a)> = I_{Max}^2 \frac{H_a^2}{H_J^2}\left(\frac{1}{1+\frac{H_a^2}{H_J^2}}\right) \quad (35)$$

where $I_{Max} = \frac{45}{64}\pi^2 k^2 a^2 j_1$, $H_J = \frac{15}{16}\frac{\sqrt{3}\phi_0}{\pi^{3/2}kad}$, $\phi = kadH$, and $I_1 = \pi k^2 a^2 j_1$. In terms of magnetic flux ($\phi$) Eq. (34) reduces to (Fig (5)-a)

$$<I_{Rec}^2\left(\frac{\phi}{\phi_0}\right)> = \frac{I_1^2}{16\pi}\frac{\phi_0^2}{\phi^2}\left\{\frac{9}{4}-\left[3-3\pi^3\frac{\phi^2}{\phi_0^2}+\frac{\pi^6}{4}\frac{\phi^4}{\phi_0^4}\right]e^{-\frac{\pi^3\phi^2}{4\phi_0^2}}+\left[\frac{3}{4}-3\pi^3\frac{\phi^2}{\phi_0^2}+\pi^6\frac{\phi^4}{\phi_0^4}\right]e^{-\pi^3\frac{\phi^2}{\phi_0^2}}\right\} \quad (36)$$

To calculate the critical current of grains with circular d-d junction model, we consider the approximation of the Struve function as [41].

$$H_1(x) = \frac{2}{\pi} - K_0(x) + \left(\frac{16}{\pi}-5\right)\frac{\sin x}{x} + \left(12-\frac{36}{\pi}\right)\frac{1-\cos x}{x^2} \quad (37)$$

here, $K_0(x)$ is the Bessel Function [42]. By using Eq. (23) the configurational averaging in Eqs. (33) gives (Fig (5)-b)

$$<I_{Cir}^2\left(\frac{\phi}{\phi_0}\right)> = \frac{8I_1^2}{\pi^3}\frac{\phi_0^2}{\phi^2}\Bigg\{[3+(-8-24\pi+9\pi^2)\left(\frac{16}{\pi^3}\frac{\phi_0^2}{\phi^2}\right)+216(-3+\pi)^2\left(\frac{256}{\pi^6}\frac{\phi_0^4}{\phi^4}\right)]+$$

$$e^{\left(\frac{-\pi^3\phi^2}{16\phi_0^2}\right)}\left[(128-96\pi+18\pi^2)+\left(\frac{16}{\pi^3}\frac{\phi_0^2}{\phi^2}\right)(512-360\pi+63\pi^2)+\left(\frac{256}{\pi^6}\frac{\phi_0^4}{\phi^4}\right)(648-423\pi+72\pi^2)\right]+e^{\left(\frac{-\pi^3\phi^2}{64\phi_0^2}\right)}\left[\left(\frac{\pi^3\phi^2}{16\phi_0^2}\right)(-8+3\pi)+(12-6\pi)+\left(\frac{16}{\pi^3}\frac{\phi_0^2}{\phi^2}\right)(-504+384\pi-72\pi^2)+\right.$$

$$\left.\left(\frac{256}{\pi^6}\frac{\phi_0^4}{\phi^4}\right)(-2592+1728\pi-288\pi^2)\right]\Bigg\} \quad (38)$$

where $\phi = 2kadH$ and $I_1 = \pi k^2 a^2$.

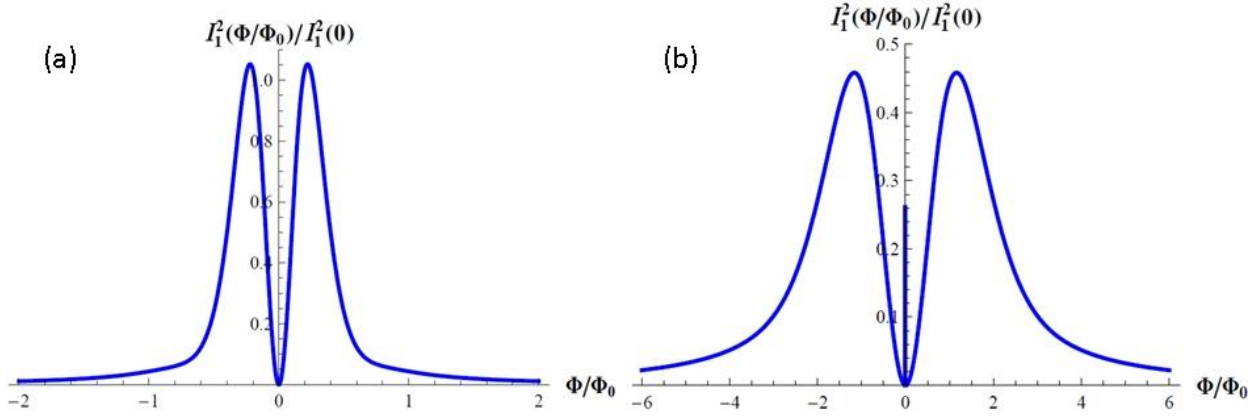

Figure 5: Mean square critical current of granular systems with d-d junction in: (a) Rectangular (b) Circular geometry

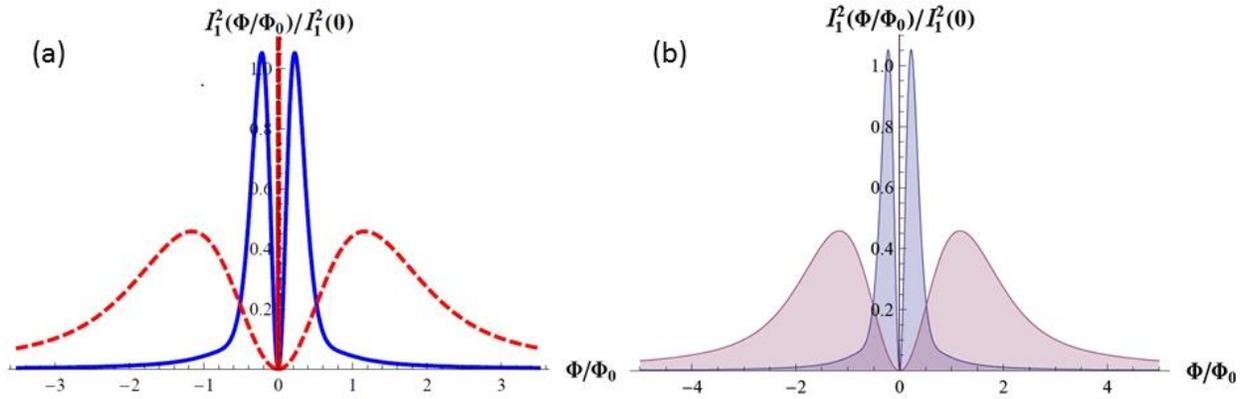

Figure 6: (a) Comparison the mean square critical current of granular system d-d junction between "rectangular model, solid line" and "circular model, dashed line" (b) show that the area surface under the curve of the circular model is much more than rectangular one.

The comparison between both model of rectangular and circular junction in the critical current of grains is shown in Fig. 6 (a). As can be seen from Fig. 6 (b) the surface area under the curve of circular geometry becomes 2.08 while for the rectangular one is 0.78. Therefore, the circular model produces better result than rectangular one for the granular critical current.

## 3-Conclusions

We have calculated the Josephson critical current of d-d junction with the twist angle $\gamma$ for one junction and a granular d-wave superconductor. We have shown that the $\gamma$ dependence for the Josephson critical current is the same for both systems.

To describe the magnetic field dependence of the critical current of a d-wave granular superconductor, in addition to the usual rectangular junction, we have also considered the circular junction model. Since the magnetic field behavior of the critical current is strongly depends on the form of the grain distribution function, for performing the averaging in Eqs. (32) and(33), we restrict ourselves to commonly Gaussian distribution law. We have shown that the circular model produces better result than rectangular one for the granular critical current.

## 4-References